\documentclass[twocolumn,prb,showpacs,prbprintnumbers,fleqn,superscriptaddress]{revtex4}
\usepackage{epsfig}
\usepackage{graphicx}
\usepackage{amsfonts}

\newcommand{\ct}{\cite}
\newcommand{\bi}{\bibitem}

\newcommand{\ga}{\gamma}
\newcommand{\si}{\sigma}

\newcommand{\non}{\nonumber}

\newcommand{\be}{\begin{equation}}
\newcommand{\ee}{\end{equation}}
\newcommand{\ba}{\begin{eqnarray}}
\newcommand{\ea}{\end{eqnarray}}

\begin{document}

\title{Defect production due to quenching through a multicritical point}
\author{Uma Divakaran}
\email{udiva@iitk.ac.in}
\author{Victor Mukherjee}
\email{victor@iitk.ac.in}
\author{Amit Dutta}
\email{dutta@iitk.ac.in}
\affiliation{Department of Physics, Indian Institute of Technology, Kanpur 
208 016, India}
\author{Diptiman Sen}
\email{diptiman@cts.iisc.ernet.in}
\affiliation{Center for High Energy Physics, Indian Institute of Science,
Bangalore 560 012, India}

\date{\today}

\begin{abstract}
We study the generation of defects when a quantum spin system is quenched 
through a multicritical point by changing a parameter of the Hamiltonian as 
$t/\tau$, where $\tau$ is the characteristic time scale of quenching.
We argue that when a quantum system is quenched across a 
multicritical point, the density of defects ($n$) in the final state is not 
necessarily given by the Kibble-Zurek scaling form $n \sim 1/\tau^{d \nu/(z 
\nu +1)}$, where $d$ is the spatial dimension, and $\nu$ and $z$ are 
respectively the correlation length and dynamical exponent associated with 
the quantum critical point. We propose a generalized scaling form of the 
defect density given by $n \sim 1/\tau^{d/(2z_2)}$, where the exponent $z_2$ 
determines the behavior of the off-diagonal term of the $2 \times 2$ 
Landau-Zener matrix at the multicritical point. This scaling is valid not 
only at a multicritical point but also at an ordinary critical point.
\end{abstract}

\pacs{73.43.Nq, 05.70.Jk, 64.60.Ht, 75.10.Jm}
\maketitle

\section{Introduction}

The zero temperature quantum phase transitions occurring in quantum many
body systems has been a challenging area of research for the past few 
years \ct{sachdev99,dutta96}. The dynamics taking place in such systems by 
varying a parameter in the Hamiltonian in a definite fashion have come to the 
forefront only recently \ct{zurek,polkovnikov05,levitov,sengupta04,bose}.
In this paper, we focus on the density of defects generated
when a system, prepared in its ground state, is adiabatically quenched at a 
uniform rate \ct{zurek,polkovnikov05,levitov,mukherjee07,divakaran07,
sen,dziarmaga06,dziarmaga08,polkovnikov08,polkovnikov07,sengupta08,santoro08,
patane08,mukherjee08,divakaran08,caneva08,cucchietti07}. These works
have their roots originating from the study of phase transitions in the early
universe \ct{kibble76} which was extended to second order phase transitions
\ct{zurek96} and later to quantum spin chains\ct{zurek}.
The diverging relaxation time associated with a quantum critical point 
results in the failure of the system to follow its instantaneous ground state;
this eventually leads to the generation of defects in the final state. When a
parameter of the quantum Hamiltonian is varied as $t/\tau$, where $\tau$ is 
the characteristic time scale of the quenching, the Kibble-Zurek (KZ) argument
\ct{zurek,polkovnikov05} predicts a density of defects in the final 
state that scales as $1/\tau^{d\nu/(z \nu+1)}$ in the limit $\tau \to 
\infty$. Here $\nu$ and $z$ denote the correlation length and dynamical 
exponents, respectively, characterizing the associated quantum phase 
transition of the $d$-dimensional quantum system. The KZ prediction has been 
verified for various exactly solvable spin models when quenched across a 
critical point \ct{zurek,levitov,mukherjee07,divakaran07}. Various 
generalizations of the KZ scaling form have also been proposed for quenching 
through a gapless phase or along a gapless line \ct{sen,santoro08,divakaran08}.
Experimental verification of the dynamics of such systems can be realized by 
the trapped ultra cold atoms in optical lattices; for a review see 
Ref. \onlinecite{bloch07}.

The generation of defects during the adiabatic quenching dynamics of a 
one-dimensional spin-1/2 $XY$ chain across a quantum critical point was 
studied in ref.\onlinecite{levitov}. The Hamiltonian of the system is 
\ct{lieb61}
\be H ~=~ - \frac{1}{2} ~\sum_n ~(J_x \si^x_n \si^x_{n+1} ~+~ J_y \si^y_n 
\si^y_{n+1} + h \si^z_n), \label{h1} \ee
where the $\si$'s are Pauli spin matrices satisfying the usual commutation 
relations. The strength of the transverse field is denoted by $h$, and $J_x$ 
and $J_y$ are the strength of the interactions in the $x$ and $y$ directions,
respectively. The phase digram of the above model is shown in Fig. 1.

\begin{figure}[htb]
\includegraphics[height=3in,width=3.4in]{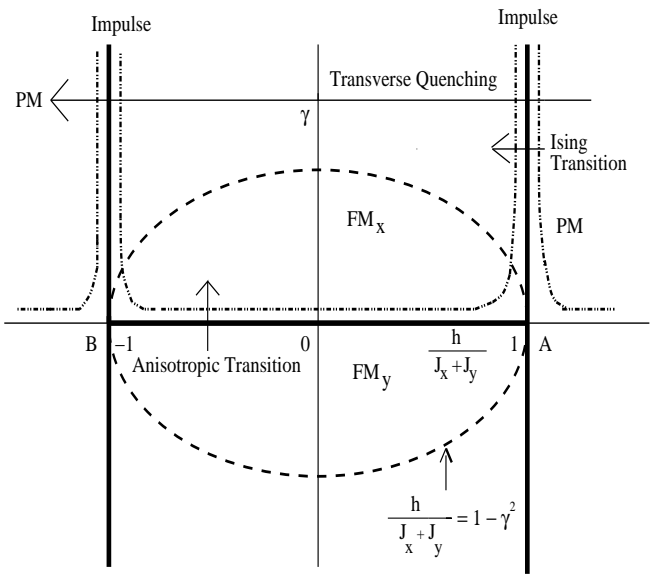}
\caption{The phase diagram of the anisotropic $XY$ model in a transverse field
in the $~h/(J_x + J_y) ~- ~\ga ~$ plane, where $\gamma \equiv (J_x -J_y)/(J_x 
+ J_y)$. The vertical bold lines given by $~h/(J_x + J_y) = \pm 1~$ denote the
Ising transitions. The system is also gapless on the horizontal bold line $\ga
= 0$ for $~|h| < J_x + J_y$. ${\rm FM_x ~(FM_y)}$ is a long-range ordered 
phase with ferromagnetic ordering in the $x ~(y)$ direction. The thick dashed 
line marks the boundary between the commensurate and incommensurate 
ferromagnetic phases. The thin dotted lines indicate the adiabatic and impulse
regions when the field $h$ is quenched from $-\infty$ to $\infty$. The two 
points with coordinates $\ga = 0$ and $h/(J_x+J_y) =\pm 1$ denoted by $A$ 
and $B$ are multicritical points.} \end{figure}

It was observed that when the transverse field $h$ is varied as $h =t/\tau$, 
the system crosses the Ising transition lines as shown in the figure, and the 
defect density scales as\ct{levitov} $1/\sqrt {\tau}$. This is in agreement 
with the KZ prediction since the values of the critical exponents associated 
with the Ising transition are given by $z=\nu=1$. On the other hand, if the 
interaction in the $x$-direction ($J_x$) is quenched in a similar fashion 
keeping $h$ and $J_y$ fixed\ct{mukherjee07}, the defect density is again found
to scale as $1/\sqrt{\tau}$ though the magnitude depends upon the values of 
$J_y$ and $h$. If $h<2J_y$, the
system crosses the anisotropic critical line ($J_x=J_y$) in addition to the 
Ising transition lines mentioned above, and hence the magnitude of the defects
is increased. However, it was observed that if $J_x$ is quenched keeping $h=
2J_y$, the system crosses the multicritical point at $J_x=J_y$ and $h=2J_y$, 
where the Ising and anisotropic transition lines meet. The density of defects 
in the final state generated in a passage through the above multicritical 
point shows a slower decay with $\tau$ given as $1/\tau^{1/6}$. Since 
the critical exponents associated with this multicritical point are 
given by $\nu=1/2$ and $z=2$, the above scaling relation does {\it not} 
follow from the KZ scaling relation $1/\tau^{d\nu/(z \nu +1)}$. It
is this observation which motivated us to look for a generalized
scaling relation valid even for a multicritical point. It should be noted
here that this is the first attempt to provide a generalized scaling 
relation for defect density when the system is quenched linearly through a 
multicritical point which has also been extended to the nonlinear case in a
recent work
\ct{sengupta082} .

The paper is organized as follows. In Sec. II, we derive the general form
for the scaling of defects and apply it in two models. Sec. III consists of 
concluding remarks.

\section{General Scaling}

To propose a general scaling scheme valid even for a multicritical point using
the Landau-Zener non-adiabatic transition probability \ct{landau,sei}, let us 
consider a $d$-dimensional model Hamiltonian of the form
\ba H &=& \sum_{\vec k} ~\psi^{\dagger}(\vec k) ~[~ \left( \lambda (t)
+ b(\vec k) ~\right) \si^z \non \\
& & ~~~~~~~~~~~~~~+~ \Delta(\vec k) ~\si^+ ~+~ \Delta^* (\vec k) ~\si^- ~]~ 
\psi(\vec k), \label{ham} \ea
where $\si^{\pm} = (\si^x \pm i \si^y)$, $b(\vec k)$ and $\Delta (\vec k)$ 
are model dependent functions, and $\psi(\vec k)$ denotes the fermionic 
operators $(\tilde \psi_1 (\vec k), \tilde \psi_2 (\vec k))$. The above 
Hamiltonian can represent, for example, a one-dimensional transverse Ising or 
$XY$ spin chain \ct{lieb61}, or an extended Kitaev model in $d=2$ written in 
terms of Jordan-Wigner fermions \ct{kitaev}. We assume that the parameter 
$\lambda(t)$ varies linearly as $t/\tau$ and vanishes at the quantum 
critical point at $t=0$, so that the system crosses a gapless point at $t=0$ 
for the wave vector $\vec k =\vec k_0$. Without loss of generality, we set
$|\vec k_0|=0$. The parameters $b(\vec k)$ and $\Delta (\vec k)$ are assumed 
to vanish at the quantum critical point in a power-law fashion given by
\be b(\vec k)\sim |\vec k|^{z_1} ~~~{\rm and}~~~ \Delta(\vec k)\sim |\vec 
k|^{z_2}. \label{exp} \ee
Many of the models described by Eq. (\ref{ham}) exhibit a quantum phase 
transition with the exponents associated with the quantum critical point being
$\nu=z=z_2=1$. We shall however explore the more general case below.

The Schr\"odinger equation describing the time evolution of the system when 
$\lambda$ is quenched is given by $i \partial \psi /\partial t = H \psi$ 
(where we set Planck's constant $\hbar =1$). Using the Hamiltonian in Eq. 
(\ref{ham}), we can write
\ba i \frac {\partial \tilde \psi_1(\vec k)}{\partial t} &=& \left(\frac t 
{\tau} + b(\vec k)\right) ~\tilde \psi_1(\vec k) ~+~ \Delta(\vec k) ~\tilde
\psi_2 (\vec k), \non \\
i \frac {\partial \tilde \psi_2(\vec k)}{\partial t} &=& -\left( \frac t 
{\tau} + b(\vec k)\right) ~\tilde \psi_2 (\vec k) ~+~ \Delta^*(\vec k)~ 
\tilde \psi_1(\vec k). \ea
One can now remove $b(\vec k)$ from the above equations by redefining 
$t/\tau + b(\vec k) \to t$; thus the exponent $z_1$ defined in Eq. (\ref{exp})
does not play any role in the following calculations. Defining a new set of 
variables 
$\psi_1 (\vec k) = \tilde \psi_1 (\vec k) \exp (i \int^t dt' ~t'/\tau)$ and
$\psi_2 (\vec k) = \tilde \psi_2 (\vec k) \exp (-i \int^t dt' ~t'/\tau)$, 
we arrive at a time evolution equation for $\psi_1 (\vec k)$ given by
\be \left(\frac {d^2}{dt^2} ~-~ 2i \frac {t}{\tau} \frac {d}{dt} ~+~ |\Delta
(\vec k)|^2 \right)\psi_1(\vec k) ~=~ 0. \ee
Further rescaling $t \to t\tau^{1/2}$ leads to 
\be \left(\frac {d^2}{dt^2} ~-~ 2i {t} \frac {d}{dt} ~+~ |\Delta(\vec k)|^2
\tau \right)\psi_1(\vec k) ~=~ 0. \ee
If the system is prepared in its ground state at the beginning of the 
quenching, i.e., $\psi_1(\vec k)=1$ at $t=-\infty$, the above equation 
suggests that the probability of the non-adiabatic transition, $p_{\vec k} =
\lim_{t \to +\infty} |\psi_1(\vec k)|^2$, must have a functional dependence on 
$|\Delta(\vec k)|^2\tau$ of the form
\be p_{\vec k} ~=~ f(|\Delta(\vec k)|^2\tau). \ee
The analytical form of the function $f$ is given by the general Landau-Zener 
formula \ct{landau,sei}. The defect density in the final state is therefore 
given by
\be n ~=~ \int\frac{d^d k}{(2\pi)^d} ~f(|\Delta(\vec k)|^2\tau) ~=~ \int
\frac{d^d k} {(2\pi)^d} ~f(|\vec k|^{2z_2}\tau). \ee
The scaling $k \to k^{2z_2}\tau$ finally leads to a scaling of the defect 
density given by 
\be n ~\sim~ 1/\tau^{d/(2z_2)}. \label{def} \ee

We shall recall the example of the quenching dynamics of the transverse $XY$ 
spin chain when the field or the interaction is quenched 
\ct{levitov,mukherjee07}. When the system is quenched across the Ising or 
anisotropic critical line by linearly changing $h$ or $J_x$ as $t/\tau$, 
$\Delta (\vec k)$ vanishes at the critical point as $\Delta (\vec k) \sim |\vec
k|$ yielding $z_2=z=1$; hence the generalized scaling form given in Eq.
(\ref{def}) matches with the Kibble-Zurek prediction with $\nu = z=1$. On the 
other hand, when the system is swept across the multicritical point $(J_x=J_y,
h=2J_y)$ by quenching the interaction $J_x=t/\tau$ with $h=2J_y$, the 
equivalent $2 \times 2$ Hamiltonian matrix of the Jordan-Wigner fermions in 
an appropriate basis can be written as \ct{mukherjee07}
\ba \left[ \begin{array}{cc} J_x + J_y (\cos 2k + 2 \cos k) & J_y (\sin 2k
+ 2 \sin k) \\
J_y (\sin 2k + 2 \sin k) & -J_x - J_y (\cos 2k + 2 \cos k) \end{array} \right].
\non \ea

The corresponding Schr\"odinger equations are
\ba i \frac {\partial \tilde \psi_1(\vec k)}{\partial t} &=& \left(\frac t 
{\tau} + J_y(\cos2k+2\cos k)\right) ~\tilde \psi_1(\vec k) \nonumber\\
&+& J_y(\sin 2k+2\sin k) ~\tilde
\psi_2 (\vec k), \non \\
i \frac {\partial \tilde \psi_2(\vec k)}{\partial t} &=& J_y(\sin 2k +2 
\sin k) ~\tilde \psi_1 (\vec k)\non \\ 
&-& \left(\frac t{\tau} + J_y(\cos 2k+2\cos k)\right)~ \tilde \psi_2(\vec k)
\label{multieq} \ea
At the quantum critical point $J_x=J_y$, the diagonal term $~b(k) = J_y
(\cos 2k + 2 \cos k)~$ goes as $~-J_y - J_y |\pi - k|^2~$ near $k=\pi$.
Hence the dynamical exponent is given by $z=z_1=2$ at this multicritical 
point. Note that in this example, the critical point is not crossed at $t=0$; 
however, one can shift the time so that $~b^{'}(k) \sim |\pi - k|^{z_1}$, 
which would ensure that the quantum critical point is crossed at $t=0$.

On the other hand, the off-diagonal term $~\Delta (k) = J_y (\sin 2k + 2 \sin 
k) =|\pi - k|^3~$ leads to the density of defect scaling as $1/\tau^{1/6}$; 
this is in agreement with the generalized scaling relation proposed in Eq. 
(\ref{def}) with $z_2=3$. Fig. 2 shows the numerical integration of Eq. 
(\ref{multieq}) which confirms the defect scaling exponent of $-1/6$.

\begin{figure}
\includegraphics[height=2.6in,width=3.4in]{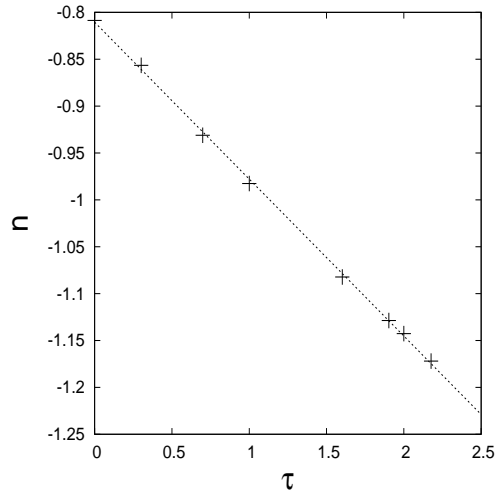}
\caption{$n$ vs $\tau$ obtained by numerically solving Eq. (\ref{multieq}) at 
the multicritical point with $h=10$ and $J_y=5$. The line has a slope of 
$-0.16$.} \end{figure}

Finally, let us comment on the dynamics of an exactly solvable transverse Ising
model with an additional three-spin interaction which is also quenched through
a multicritical point\ct{divakaran07} by varying the transverse field as 
$h=t/\tau$. It has been observed that the defect density $n$ scales as 
$1/\tau^{1/6}$ which again does not support the KZ scaling form.
The three-spin interacting Hamiltonian is given by \ct{kopp05}
\ba H = - \frac{1}{2} \sum_i ~\si^z_i [h + J_3 \si^x_{i-1} \si^x_{i+1}] +
\frac{J_x}{2} \sum_i ~\si^x_i \si^x_{i+1}. \ea
We shall henceforth set $J_x=1$. The equivalent $2 \times 2$ Hamiltonian 
matrix of the Jordan-Wigner fermions in the momentum representation takes the
form 
\[ \left[ \begin{array}{cc} h(t)+\cos k-J_3\cos 2k & i (\sin k-J_3\sin 2k) \\
-i(\sin k-J_3\sin 2k) & -(h(t)+\cos k-J_3\cos 2k) \end{array} \right] \]
It may be noted that by virtue of a duality transformation, this model can be 
mapped to a transverse $XY$ model with competing interactions for the $x$ and
$y$ components of the spin \ct{kopp05}. The multicritical point in the phase 
diagram of this model is at $h=-1$ and $J_3=1/2$. We observe that the 
off-diagonal term $~\sin k-J_3\sin 2k ~$ scales as $|\vec k|^3$ at the 
multicritical point; therefore the defect density scales as $1/\tau^{1/6}$ as 
expected from the general scaling relation proposed here. The importance of 
the multicritical point has also been observed in nonlinear quenching of 
different models \ct{sengupta082}.

\section{Conclusions}

We have shown that the density of defects $n$ produced when a system is 
quenched through a multicritical point does not follow the KZ scaling relation
$1/\tau^{d\nu/(z \nu +1)}$. We have then proved a new scaling form which is 
not only valid at an ordinary quantum critical point but is also valid at 
a multicritical point. We argue that for a system which is swept across a 
multicritical point in a phase diagram, it is the exponent $z_2$ defined 
above which appears in the scaling of the defect density given in Eq. 
(\ref{def}). However, for a passage through an ordinary critical point in many
models, $z_2=z=1$, and Eq. (\ref{def}) reproduces the conventional KZ scaling 
form with $\nu=z=1$.

\begin{center}
\bf Acknowledgments
\end{center}

AD acknowledges Subir Sachdev for a very interesting discussion. AD and UD 
thank G. E. Santoro for stimulating discussions and the hospitality of SISSA, 
Trieste, Italy where part of this work was carried out. AD and DS also thank 
Krishnendu Sengupta for his comments. DS thanks DST, India for financial 
support under Project No. SR/S2/CMP-27/2006.

\end{document}